\pdfoutput=1

\documentclass[11pt]{article}
\usepackage{acl}

\usepackage{times}
\usepackage{latexsym}

\usepackage[T1]{fontenc}

\usepackage[utf8]{inputenc}

\usepackage{color}

\usepackage[table]{xcolor}
\usepackage[normalem]{ulem}  
\usepackage{colortbl}   
\usepackage{contour}
\usepackage{ulem}

\contourlength{0.8pt}

\definecolor{searchhigh}{RGB}{255,180,180}
\definecolor{searchlow}{RGB}{255,230,230}
\definecolor{refhigh}{RGB}{180,255,180}
\definecolor{reflow}{RGB}{230,255,230}
\usepackage{arydshln}

\usepackage{tabularx}
\usepackage{tabu}
\usepackage{booktabs}
\usepackage{multirow}

\usepackage{graphicx} 
\usepackage{float}
\usepackage{subfigure} 
\usepackage{amssymb}

\usepackage{hyperref}

\usepackage{url}
\usepackage{tablefootnote}
\usepackage{xspace}

\usepackage{float}
\usepackage{amsmath}
\usepackage{cleveref}
\crefname{tcolorbox}{box}{boxes}
\Crefname{tcolorbox}{Box}{Boxes}
\usepackage{bm}
\usepackage{algorithmicx}
\usepackage{algorithm}
\usepackage{algpseudocode}
\usepackage{placeins}
\usepackage{arydshln}

\usepackage{amssymb}
\usepackage{pifont}
\usepackage{enumitem}

\usepackage{xspace}
\usepackage[most]{tcolorbox}
\newtcolorbox{blueBox}[1][]{
  colback=blue!5!white,
  colframe=blue!75!black,
  title=#1,
  fonttitle=\bfseries,
  boxrule=0.7pt,
  arc=4pt,
  left=3pt,
  right=3pt,
  top=3pt,
  bottom=3pt
}

\newtcolorbox{wronganswer}[1][]{
    enhanced,
    breakable,
    colframe=customred,
    colback=customred!10!white,
    sharp corners,
    boxsep=0pt,
    left=5pt,
    right=5pt,
    top=6pt,
    bottom=6pt,
    boxrule=0pt,
    leftrule=4pt,
    #1
}
\definecolor{customred}{RGB}{255,0,0} 
\usepackage{microtype}

\usepackage{inconsolata}
\newcommand{\ours}{\textsc{Sage}\xspace}
\newcommand{\eg}{\hbox{\emph{e.g.,}}\xspace}
\newcommand{\ie}{\hbox{\emph{i.e.,}}\xspace}

\newcommand{\github}{\raisebox{-1.5pt}{\includegraphics[height=1.05em]{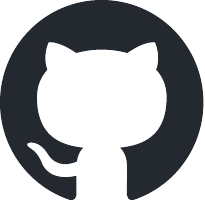}}\xspace}
\renewcommand{\thefootnote}{\arabic{footnote}}

\addtocontents{toc}{\protect\setcounter{tocdepth}{-1}}
\title{\ours: Benchmarking and Improving Retrieval for Deep Research Agents}

\newcommand{\NYU}{$^{1}$}
\newcommand{\YALE}{$^{2}$}
\newcommand{\CDS}{$^{3}$}
\newcommand{\drtulu}{DR Tulu\xspace}

\definecolor{YaleYellow}{RGB}{179, 176, 4} 
\definecolor{NYUPurple}{RGB}{134, 1, 175}  
\definecolor{NTUBlue}{RGB}{2,2,200} 
\definecolor{Alibaba}{RGB}{255, 106, 0}
\definecolor{Center}{RGB}{0, 128, 0}

\author{
  \textbf{Tiansheng~Hu}\NYU \quad
  \textbf{Yilun~Zhao}\YALE \quad
  \textbf{Canyu~Zhang}\NYU \quad
  \textbf{Arman~Cohan}\YALE \quad
  \textbf{Chen~Zhao}\NYU $^,$\CDS $^\dagger$\\[4pt]
  \NYU\,NYU Shanghai \quad
  \YALE\,Yale University \quad 
  \CDS\,Center for Data Science, New York University \\[4pt]
  \github ~~~\url{https://github.com/HughieHu/Sage}
}

\begin{document}
\maketitle
\begin{abstract}
Deep research agents have emerged as powerful systems for addressing complex queries. Meanwhile, LLM-based retrievers have demonstrated strong capability in following instructions or reasoning. This raises a critical question: can LLM-based retrievers effectively contribute to deep research agent workflows? To investigate this, we introduce \ours, a benchmark for scientific literature retrieval comprising 1,200 queries across four scientific domains, with a 200,000 paper retrieval corpus.
We evaluate six deep research agents and find that all systems struggle with reasoning-intensive retrieval. Using \drtulu as backbone, we further compare BM25 and LLM-based retrievers (\ie ReasonIR and gte-Qwen2-7B-instruct) as alternative search tools. Surprisingly, BM25 significantly outperforms LLM-based retrievers by approximately 30\%, as existing agents generate keyword-oriented sub-queries. 
To improve performance, we propose a corpus-level test-time scaling framework that uses LLMs to augment documents with metadata and keywords, making retrieval easier for off-the-shelf retrievers. This yields 8\% and 2\% gains on short-form and open-ended questions, respectively.

\renewcommand{\thefootnote}{\fnsymbol{footnote}}
\setcounter{footnote}{0}
\footnotetext[2]{Correspondence: Chen Zhao (\texttt{cz1285@nyu.edu})}
\renewcommand{\thefootnote}{\arabic{footnote}}

\end{abstract}

\section{Introduction}
\begin{figure}[t]
    \centering
    \includegraphics[width=\columnwidth, height=0.4\textheight, keepaspectratio]{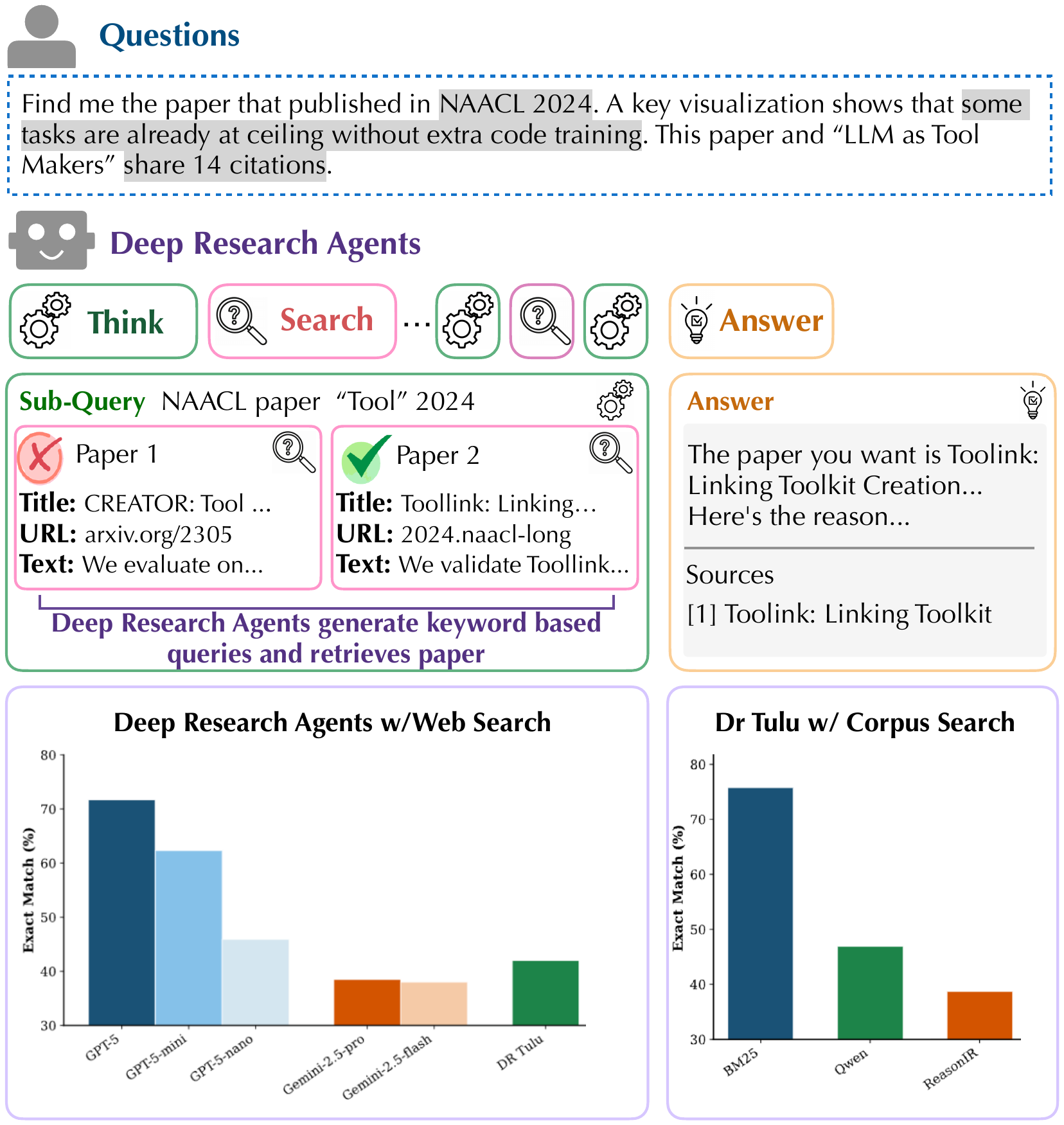}
    \caption{\ours task overview. Given a complex question, the deep research agent (\eg \drtulu) iteratively reasons, generates keyword-based sub-queries, searches for relevant papers, and outputs a final answer. We first evaluate the agents with their native web-search tool, and then modify \drtulu's MCP service to replace web search with retrievers that performs corpus search over our paper collection.}
    \label{fig:Task}
\end{figure}

Like human experts, deep research agents \citep{openai2025deepresearch,gemini_2025_deep_reserach,perplexity_2025_deep_reserach,shao2025drtulureinforcementlearning} address complex queries by iteratively searching and synthesizing information across multiple sources. With the help of recent advances in the agentic capabilities of large language models (LLMs), these systems demonstrate strong and robust performance in benchmarks across multiple domains \citep{agashe2025agent,zheng-etal-2025-deepresearcher,li-etal-2025-quantagents,wang2025lmarslegalmultiagentworkflow,chervonyi2025goldmedalistperformancesolvingolympiad,zhao2025sciarena}.

At the core of deep research agents lie their retrieval stack \citep{zheng-etal-2025-deepresearcher,besrour2025ragentamultiagentretrievalaugmentedgeneration}. Recent advances in LLM-based retrievers have shown strong promise, particularly in their ability to follow instructions and support reasoning-intensive retrieval \citep{shao2025reasonir,muennighoff2025generative,weller2025promptriever}. However, most existing commercial deep research agents adopt proprietary search APIs over large web corpora, which rely on surface-form matching. We thus ask the following research question: \textbf{Whether LLM-based retrievers can effectively contribute to deep research agent workflows?}

We propose to systematically study the retrieval behaviors of deep research agents with a scientific literature search task. As shown in Figure~\ref{fig:Task}, queries in this task often require a deep understanding of research concepts as well as the ability to reason across entire scholarly articles. Moreover, unlike open-domain web search, this task provides a controllable experimental environment with a fixed and well-defined corpus of scientific papers. To this end, we introduce \ours,
a deep-research benchmark for \textbf{S}cientific \textbf{AG}entic retrieval \textbf{E}valuation, consisting of 1,200 queries over a corpus of 200,000 papers spanning four scientific domains. \ours includes two complementary types of questions: (1) \emph{short-form} questions with a verifiable answer that often require intensive reasoning, and (2) \emph{open-ended} questions that reflect practical research tasks such as searching related work.  

We first evaluate six deep research agents, including both proprietary systems like GPT-5 \citep{gpt_5} and Gemini-2.5-Pro \citep{gemini_2025_pro} and the open-source one \drtulu \citep{shao2025drtulureinforcementlearning}. While proprietary agents perform best and \drtulu is competitive, all systems struggle with reasoning-intensive retrieval that requires synthesizing metadata and inter-paper relationships. Using \drtulu as the backbone agent, we further find that BM25 \citep{robertson-1994-okapi} significantly outperforms LLM-based retrievers by about 30\%. Analysis shows that the sub-queries generated by existing deep research agents are keyword-oriented. This behavior aligns well with surface-form matching, while the semantic capabilities of LLM-based retrievers falter due to mismatched query formulations.

To address the reasoning-intensive retrieval challenge, we propose a novel corpus-level test-time scaling framework. The key idea is to leverage LLMs to reason over each paper and enrich the corpus with additional signals that make retrieval easier for off-the-shelf retrievers. Specifically, we augment each paper with informative metadata and keywords. This approach yields substantial improvements on \ours, achieving  8\% gains on short-form questions and 2\% on open-ended questions.

We summarize our key contributions as follows:

\begin{itemize}[leftmargin=*,noitemsep,topsep=2pt]
    \item We introduce \ours, a reasoning intensive benchmark combining short-form queries and open-ended queries together with a large dataset.
    \item We conduct extensive evaluation and find that LLM-based retrievers collaborate poorly with deep-research agent.
    \item We introduce a new framework for corpus-level test-time scaling and achieve great improvements on both short-form and open-ended queries.
\end{itemize}

\section{Related Work}
\label{sec:related}

\paragraph{Deep Research Agents.}
Deep research agents represent a new paradigm of autonomous AI systems designed to tackle complex, multi-step information-seeking tasks~\citep{huang2025deepresearch}.
Commercial systems including OpenAI's Deep Research~\citep{openai2025deepresearch}, Google's Gemini Deep Research~\citep{gemini_2025_flash} have demonstrated impressive performance on challenging benchmarks such as BrowseComp~\cite{wei2025browsecompsimplechallengingbenchmark}. In parallel, open-source efforts have rapidly advanced, with systems such as SearchR1, WebThinker, and Tongyi Deep Research approaching competitive performance~\citep{li-etal-2025-search, jin2025searchr, li2025webthinkerempoweringlargereasoning, tongyideepresearchteam2025tongyideepresearchtechnicalreport}. Notably, \drtulu~\citep{shao2025drtulureinforcementlearning} is the first open model explicitly trained for open-ended, long-form deep research via reinforcement learning, achieving results comparable to proprietary systems on benchmarks.
Despite these advances, existing deep research agents rely primarily on web search or proprietary retrieval backends. Whether such agents can function as plug-and-play solutions when paired with LLM-based retrievers over closed-domain corpora remains largely unexplored, which we systematically investigate in this work.

\paragraph{LLM-based Retrievers.}
The advent of large-scale contrastive learning marked a significant advancement for retrievers~\cite{ni-etal-2022-large, gao-etal-2023-precise, li2023towards, wang2024multilinguale5textembeddings, chen-etal-2024-m3}. 
More recently, decoder-based retrievers such as LLM2Vec~\citep{behnamghader2024llmvec} and GritLM~\citep{muennighoff2025generative} have emerged, repurposing generative LLMs for embedding tasks.
Beyond general-purpose embeddings, recent work has explored training LLM-based retrievers to enhance specific capabilities. Promptriever~\citep{weller2025promptriever} introduces instruction-trained retrievers that can be prompted like language models. ReasonIR~\citep{shao2025reasonir} presents the first retriever specifically trained for reasoning-intensive tasks such as finding similar coding problems. However, whether these retrievers can collaborate effectively with agentic search paradigms remains unexplored, and our work bridges this gap.

\paragraph{Test-time Scaling for Retrieval.}

Test-time scaling has emerged as an effective paradigm for enhancing model performance by allocating additional computation during inference~\citep{snell2025scaling, muennighoff-etal-2025-s1}. Within retrieval domain, Rank1~\citep{weller2025rank} introduces the first reranking model trained to leverage test-time compute. Other approaches explore query expansion~\citep{gao-etal-2023-precise}, query rewriting~\citep{ma-etal-2023-query} and to further leverage inference-time computation. Here in our work, we investigate how corpus-level test-time scaling can adapt the corpus to better align with automatically decomposed sub-queries from deep-research agent like \drtulu \citep{shao2025drtulureinforcementlearning} for task-specific retrieval.

\section{\ours Benchmark}
\label{sec:benchmark}

This section introduces \ours Benchmark. We begin with motivating \ours (\S\ref{sec:data}), then present data curation and evaluation metric for shot-form questions (\S\ref{sec:short}), followed by those for open-ended questions (\S\ref{sec:open-end}) and corpus construction (\S\ref{sec:corpus}).

\paragraph{Problem Formulation.} Unlike traditional RAG system, which given a query $q$, retrieves documents $\mathcal{D} = \text{Retrieve}(q)$ and generates a response conditioned on $\mathcal{D}$ in one shot, a deep research agent is an agentic system composed of one or more LLMs augmented with search tools. Such agents autonomously plan multi-step research procedures, retrieve information from online sources, and synthesize evidence into a comprehensive, well-cited answer. Specifically, the agent selects an action $a_i \in \{\texttt{think}, \texttt{tool}, \texttt{answer}\}$ at each step: reasoning internally, issuing a sub-query $q_i$ to retrieve documents $\mathcal{D}_i = \text{Retrieve}(q_i)$, or producing the final answer conditioned on the accumulated evidence $\bigcup_{j} \mathcal{D}_j$. This formulation enables the agent to decompose complex questions into sub-queries $\{q_1, q_2, \ldots, q_n\}$, progressively building an evidence base across multiple retrieval rounds.

\subsection{Why Scientific Literature Search?}
\label{sec:data}

Our primary goal is to study the retrieval behavior of deep research agents. To achieve this, we choose scientific literature search as our testbed for several reasons: (1) \textbf{Task is Common and Impactful}. Searching for relevant literature is an integral part of the research process, whether it is to verify if an idea has been explored before or to collect related work. Therefore a strong agentic system could significantly accelerate the scientific discovery process. (2) \textbf{Controllable Domain Specific Corpus}. Existing deep research tasks rely on entire web as a corpus, limited by the use of commercial search APIs. In contrast, scientific literature search adopts collections of papers as a controlled corpus for precise evaluation of different retrievers. (3)
\textbf{Existing Datasets Fall Short}.  While several datasets exist for scientific literature search, they fail to  evaluate deep research agents. This is because the papers used in these datasets are outdated and often include LLMs' pre-existing knowledge. However, scientific literature is a rapidly evolving field, with new papers published daily. Our dataset uses up-to-date papers to better study the retrieval behavior of deep research agents in a dynamic environment.

Based on these reasons, we construct \ours, which includes 1,200 questions spanning short-form and open-ended types. These questions cover four critical scientific domains: Computer Science, Natural Science, Healthcare, and Humanities. For each domain, we curate a corpus of 50,000 up-to-date papers. The statistics of our dataset are presented in \autoref{tab:dataset_statistics}.
\begin{table}[t]
\centering
\footnotesize
\setlength{\tabcolsep}{3pt}
\renewcommand{\arraystretch}{1.0}
\begin{tabular}{@{}l@{\hspace{5pt}}cccc@{}}
\toprule
\textbf{Property} & \textbf{Com. Sci.} & \textbf{Nat. Sci.} & \textbf{Health.} & \textbf{Human.} \\
\midrule
\multicolumn{5}{l}{\textit{Short-Form Questions}} \\
\quad Query Num & 150 & 150 & 150 & 150 \\
\quad Query Length & 201.5 & 180.3 & 187.6 & 188.3 \\[3pt]
\hdashline[3pt/1.5pt]
\noalign{\vspace{3pt}}
\quad GT Documents & 1.00 & 1.00 & 1.00 & 1.00 \\
\quad DB Size & 47,637 & 50,000 & 50,000 & 39,032 \\
\midrule
\multicolumn{5}{l}{\textit{Open-Ended Questions}} \\
\quad Query Num & 150 & 150 & 150 & 150 \\
\quad Query Length & 99.6 & 103.9 & 101.5 & 101.2 \\[3pt]
\hdashline[3pt/1.5pt]
\noalign{\vspace{3pt}}
\quad GT Documents & 17.62 & 12.67 & 10.83 & 9.94 \\
\quad DB Size & 46,756 & 48,879 & 47,745 & 37,506 \\
\bottomrule
\end{tabular}
\caption{Our Benchmark statistics. Domains: Computer Science (Com. Sci.), Natural Science (Nat. Sci.), Healthcare (Health.), and Humanities (Human.). Query length is in tokens. GT Documents = average ground truth papers per query.}
\label{tab:dataset_statistics}
\end{table}

\subsection{Short-form Questions}
\label{sec:short}

\begin{figure}[t]
    \centering
    \includegraphics[width=\columnwidth, height=0.4\textheight, keepaspectratio]{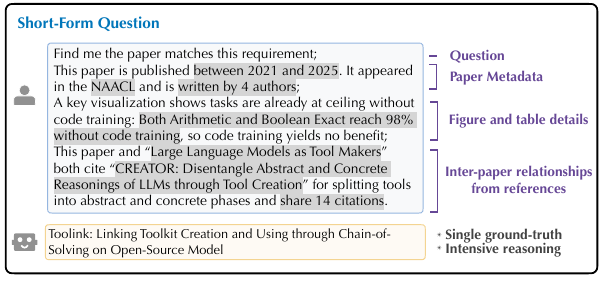}
    \caption{Overview of \emph{short-form} questions that require intensive reasoning over metadata, paper details and inter-paper relationships. Each question consists of three parts and has only one ground-truth answer.}
    \label{fig:short_form_questoin}
\end{figure}

The first type of questions in \ours is short-form questions. Similar to existing deep research benchmarks \citep{wei2025browsecompsimplechallengingbenchmark,chen2025browsecompplusfairtransparentevaluation}, short-form questions emphasize two key characteristics: (1) \textbf{Intensive Reasoning.} These questions require deep research agents to browse multiple papers, synthesize detailed and scattered information, and derive a final answer; (2) \textbf{Verifiability.} The answer to each question is unique and fixed, therefore the correctness is easily verifiable. An example of short-form question can be found at \autoref{fig:short_form_questoin}.

\paragraph{Data Curation.}
We construct question-answer pairs from three sources: extracted paper metadata (\eg author count, title length), figures and tables extracted using PyMuPDF \citep{PyMuPDF}, and inter-paper relationships established via reference overlap. To establish inter-paper relationships, we compute the citation overlap between papers, which we consider two papers as related if they share at least four common references in their reference lists. 
Specifically, we first sample a seed paper and a related paper published after 2024 from major venues in each domain (\eg ACL, ICML, NeurIPS for computer science). Next, we extract the corresponding metadata, figures, tables, and inter-paper relationships. We then prompt LLMs (GPT-5-mini \citep{gpt_5} in this case) to generate questions that require reasoning across these multiple sources. The answer to each question is the seed paper itself.

\paragraph{Evaluation Metric.} We use \textbf{Exact Match (EM)} as the metric to evaluate whether the ground truth answer is included in the output text or citations.

\subsection{Open-Ended Questions}
\label{sec:open-end}
\begin{figure}[t]
    \centering
    \includegraphics[width=\columnwidth, height=0.4\textheight, keepaspectratio]{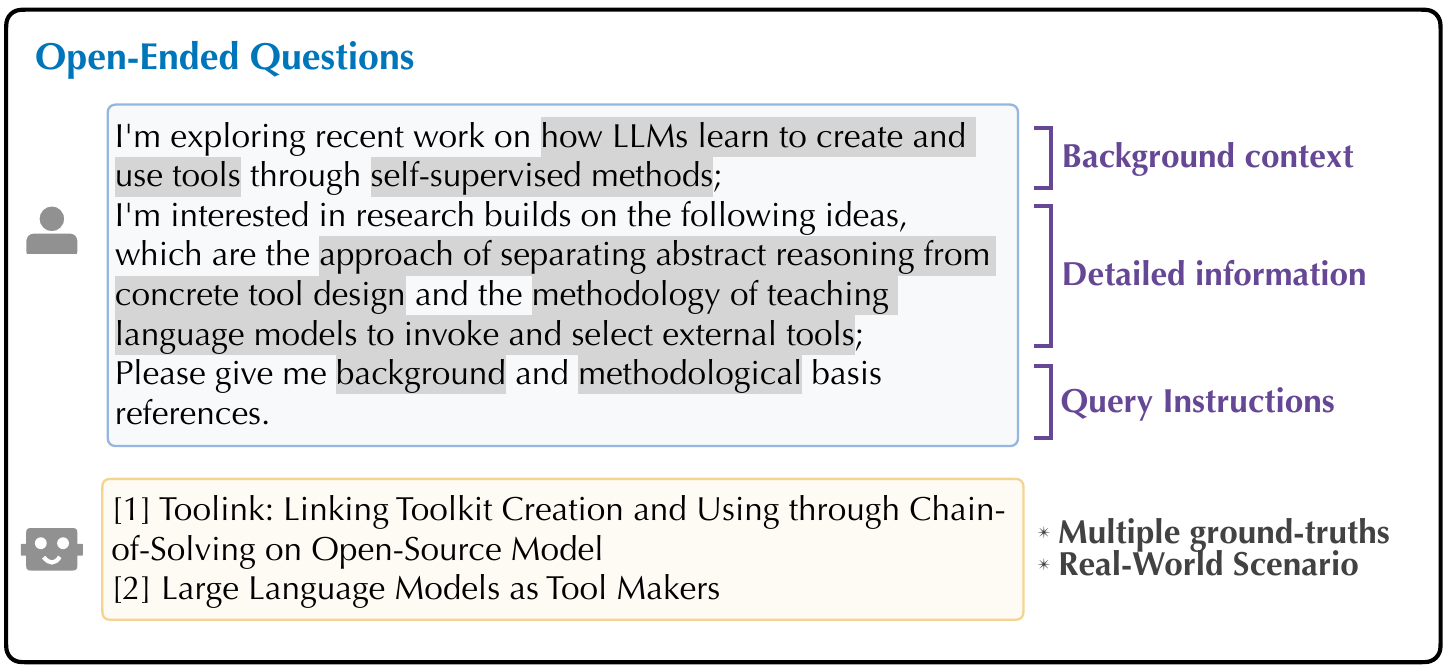}
    \caption{Overview of \emph{open-ended} questions that are grounded on real-world scenarios. Each question consists of three parts and has multiple ground-truth papers weighted by their relevance.}
    \label{fig:open_ended_questoin}
\end{figure}

Unlike short-form questions, which primarily aim to objectively measure and compare different deep research systems~\citep{rodriguez2021evaluation}, open-ended questions are grounded in real-world scenarios. They mimic the types of questions researchers encounter when conducting literature reviews and exploring new ideas. An example of open-ended question can be found at \autoref{fig:open_ended_questoin}.

\paragraph{Data Curation.}
The open-ended questions consist of two components: (1) the background context of the research topic, and (2) the shared citations between a pair of papers. We construct questions through the following pipeline: First, we leverage the reference-overlap information from Section \ref{sec:short} to select paper pairs. For each selected pair, we adopt GPT-5-mini \citep{gpt_5} to analyze the inter-relationship between the two papers and the reasons for their shared citations. Based on this analysis, GPT-5-mini \citep{gpt_5} generates corresponding questions. Note that each open-ended question has multiple ground truth papers, so we create the ground-truth using a hierarchical structure. The most relevant papers are the selected seed paper pair, followed by those cited by both papers.

\paragraph{Evaluation Metric.}
Given the list of ground-truth papers for open-ended questions, we first assign discrete relevance scores $r \in \{2, 1, 0\}$: \emph{Most Relevant} ($r{=}2$) for the two seed papers; \emph{Relevant} ($r{=}1$) for the intersection of the core papers' references; and \emph{Not Relevant} ($r{=}0$) for all others. We report \textbf{Weighted Recall} to capture all papers from both the output text and citation lists: 
\begin{equation}
\mathrm{Weighted~Recall} = \frac{\sum_{d \in L} g(\mathrm{rel}(d))}{\sum_{d \in G} g(\mathrm{rel}(d))},
\end{equation}
where $L$ is the set of retrieved documents, $G$ is the set of all relevant documents, and $g(r) = r$ is the linear gain function.

\subsection{Corpus Construction}
\label{sec:corpus}

For each domain, we construct a 50k-paper corpus using only open-access PDFs to ensure accessibility. The corpus begins with the following: (1) the ground-truth target paper and its highest-overlap partner from the computed reference-overlap information, (2) the intersection of their references, and (3) the union of their references. We then expand the corpus by sampling papers published in or after 2020 from major venues in the respective domain until the desired corpus size is reached. Due to the limited availability of papers in the humanities, this process results in approximately 40k papers, as we intentionally exclude very old literature.

\section{Experiment}
\label{sec:experiments}
\begin{table*}[t]
\centering
\small
\begin{tabular}{l cccc cc c}
\toprule
\textbf{Method} & \textbf{Com. Sci.} & \textbf{Healthcare} & \textbf{Humanities} & \textbf{Nat. Sci.} & \textbf{Avg. Searches} & \textbf{Avg. Refs} & \textbf{Avg. Perf.} \\
\midrule\midrule
\multicolumn{8}{l}{\colorbox{gray!25}{\textbf{\textsc{~~Web Search~~}}}} \\
\midrule
\multicolumn{8}{l}{\footnotesize\textbf{Short-Form Questions (Exact Match)}} \\[0.1em]
GPT-5 & \textbf{57.3} & \textbf{78.7} & \textbf{79.1} & \textbf{71.7} & 8.78 & 6.54 & \textbf{71.7} \\
GPT-5-mini & \underline{40.0} & \underline{72.0} & \underline{70.7} & \underline{66.4} & 8.15 & 4.69 & \underline{62.3} \\
GPT-5-nano & 30.7 & 46.7 & 62.0 & 44.3 & 8.92 & 7.06 & 45.9 \\
DR Tulu & 36.0 & 58.0 & 49.3 & 44.7 & \cellcolor{searchlow}7.35 & \cellcolor{refhigh}37.32 & 42.0 \\
Gemini-2.5-pro & 27.7 & 47.8 & 40.3 & 37.7 & 14.02 & \cellcolor{reflow}1.43 & 38.5 \\
Gemini-2.5-flash & 30.3 & 43.7 & 41.6 & 36.4 & \cellcolor{searchhigh}15.64 & 3.79 & 38.0 \\
\addlinespace[0.5em]
\multicolumn{8}{l}{\footnotesize\textbf{Open-Ended Questions (Weighted Recall)}} \\[0.1em]
GPT-5 & \textbf{35.1} & \underline{25.0} & \textbf{18.8} & \textbf{26.2} & \cellcolor{searchhigh}13.69 & 29.02 & \textbf{26.3} \\
GPT-5-mini & 27.4 & 17.8 & 13.9 & \underline{22.4} & 10.07 & 30.70 & 20.4 \\
GPT-5-nano & 25.9 & \textbf{29.8} & 15.7 & 21.1 & 8.59 & 30.27 & \underline{20.6} \\
DR Tulu & 18.0 & 17.2 & 14.1 & 20.2 & \cellcolor{searchlow}4.25 & \cellcolor{refhigh}35.95 & 17.4 \\
Gemini-2.5-flash & \underline{28.2} & 10.1 & \underline{15.7} & 9.8 & 6.50 & 22.12 & 16.0 \\
Gemini-2.5-pro & 20.0 & 5.2 & 12.2 & 6.7 & 10.77 & \cellcolor{reflow}14.18 & 11.0 \\
\midrule\midrule
\multicolumn{8}{l}{\colorbox{gray!25}{\textbf{\textsc{~~Corpus Search~~}}}} \\
\midrule
\multicolumn{8}{l}{\footnotesize\textbf{Short-Form Questions (Exact Match)}} \\[0.1em]
BM25 $k$=10 & \textbf{63.3} & \textbf{88.8} & \textbf{84.4} & \textbf{88.4} & 6.42 & \cellcolor{refhigh}40.0 & \textbf{81.2} \\
BM25 $k$=5 & \underline{56.0} & \underline{79.9} & \underline{82.0} & \underline{85.3} & 7.54 & 22.4 & \underline{75.8} \\
gte-Qwen $k$=10 & 44.4 & 69.7 & 73.3 & 64.4 & 5.88 & 33.2 & 63.0 \\
ReasonIR $k$=10 & 28.0 & 57.0 & 61.3 & 50.7 & 7.51 & 35.8 & 49.3 \\
gte-Qwen $k$=5 & 32.9 & 52.1 & 62.6 & 40.0 & \cellcolor{searchlow}4.82 & \cellcolor{reflow}14.2 & 46.9 \\
ReasonIR $k$=5 & 25.9 & 42.2 & 48.2 & 38.4 & \cellcolor{searchhigh}8.78 & 17.4 & 38.7 \\
\addlinespace[0.5em]
\multicolumn{8}{l}{\footnotesize\textbf{Open-Ended Questions (Weighted Recall)}} \\[0.1em]
gte-Qwen $k$=10 & \textbf{28.9} & \underline{33.5} & \textbf{36.6} & \textbf{33.0} & 4.54 & 29.3 & \textbf{33.0} \\
BM25 $k$=10 & 20.3 & \textbf{36.2} & \underline{34.0} & \underline{32.3} & \cellcolor{searchlow}4.17 & \cellcolor{refhigh}29.9 & \underline{30.7} \\
ReasonIR $k$=10 & 16.1 & 29.5 & 32.0 & 27.4 & 4.44 & 26.8 & 26.2 \\
gte-Qwen $k$=5 & \underline{22.4} & 26.3 & 29.3 & 26.2 & 4.79 & \cellcolor{reflow}15.4 & 26.0 \\
BM25 $k$=5 & 18.2 & 29.4 & 28.3 & 26.3 & 4.72 & 16.7 & 25.5 \\
ReasonIR $k$=5 & 11.3 & 18.5 & 22.3 & 17.0 & \cellcolor{searchhigh}6.21 & 15.8 & 17.3 \\
\bottomrule
\end{tabular}
\caption{Performance comparison across two question types. \textbf{Avg. Perf.} denotes the average performance across all domains. \textbf{Bold} indicates the best result and \underline{underline} indicates the second best. For \textbf{Avg. Searches}: \colorbox{searchhigh}{dark red} = highest, \colorbox{searchlow}{light red} = lowest. For \textbf{Avg. Refs}: \colorbox{refhigh}{dark green} = highest, \colorbox{reflow}{light green} = lowest.}
\label{tab:Main_Result}
\end{table*}

In this section, we first describe the experiment setup for deep research agents with web search  (\S\ref{sec:web_search_setup}) and report their results on \ours (\S\ref{sec:web_search_result}). We then move to a controlled setting by evaluating retriever performance within the same deep-research agent (\ie \drtulu) using a retrieval corpus we constructed (\S\ref{sec:corpus_search_setup} and \S\ref{sec:corpus_search_result}). At last, we presents ablation results on short-form questions (\S\ref{sec:ablation}).

\subsection{Web-Search Experiment Setup}
\label{sec:web_search_setup}
We evaluate two categories of deep research agents: (1) \textbf{Proprietary deep research agents}, including GPT-5 \citep{gpt_5}, GPT-5-mini \citep{gpt_5}, GPT-5-nano \citep{gpt_5}, Gemini-2.5-Pro \citep{gemini_2025_pro}, and Gemini-2.5-Flash \citep{gemini_2025_flash}, by using the offical APIs; (2) \textbf{Open-source deep research agents}, notably AI2's recently released \drtulu~\citep{shao2025drtulureinforcementlearning}, which sets a new SOTA among open-source deep-research agents.
For GPT series\footnote{We do not evaluate o3- and o4-mini-deep-research \citep{openai2025deepresearch}, as GPT-5 already surpasses these them on complex reasoning-intensive retrieval~\citep{gpt_5}.}, we set the \emph{``reasoning effort''} to \emph{``medium''}, and enable web search functionality. For Gemini series, we set \emph{``thinkingBudget''} to \emph{``-1''} to enable dynamic thinking and give web search permission. 
For \drtulu \citep{shao2025drtulureinforcementlearning}, we deploy the model on a server equipped with one H100 GPU and perform inference using vLLM. 

\subsection{Web-Search Results}
\label{sec:web_search_result} 
\autoref{tab:Main_Result} presents the results of deep research agents with web search. We have the following findings:

\paragraph{GPT-5 leads overall on short-form questions, while open-ended questions vary more by domain and model.} On short-form questions, the GPT-5 series delivers the strongest performance across all domains, with GPT-5 achieving the best EM (71.69\%). In contrast, open-ended questions induce more heterogeneous outcomes: GPT-5-nano performs best in healthcare, while Gemini-2.5-flash is competitive in computer science and humanities.
Notably, \drtulu outperforms the closed-source Gemini-2.5 series agents on short-form questions, indicating that open-source deep research agents can match or exceed proprietary systems in precise, retrieval-heavy settings.

\paragraph{Search quantity is not the main driver of accuracy.} On short-form questions, Gemini-2.5-flash issues nearly twice as many web-search calls as GPT-5, and \drtulu returns an exceptionally large number of references (37.32 on average), yet both trail GPT-5 by a substantial margin. This pattern suggests that brute-force searching or reference accumulation is insufficient for precise retrieval. Instead, stronger models appear to benefit from more accurate query decomposition and more targeted evidence selection, achieving higher accuracy with fewer, better-aligned searches.

\paragraph{Agents adapt search effort differently across query types.} When moving from short-form to open-ended questions, \drtulu and the Gemini series reduce the number of searches, consistent with looser constraints and potentially earlier stopping. In contrast, GPT-5 increases search activity on open-ended questions and attains the best overall results, with only a modest and acceptable increase in the number of references compared with other agents.

\paragraph{Query decomposition strategies differ across agents.} As shown in  \autoref{fig:gpt_5_decompose} and  \autoref{fig:Dr_Tulu_decompose} in Appendix, the proprietary models tend to decompose queries into more phrasal, semantically structured search queries, whereas \drtulu sub-queries more often resemble less structured keyword concatenations. This difference aligns with the observed efficiency gap, where more structured decomposition corresponds to fewer but higher-yield searches and improved retrieval precision.

\subsection{Corpus-Search Experiment Setup}
\label{sec:corpus_search_setup}
Motivated by our web-search results on the dataset, we next investigate how LLM-based retrievers integrate with deep research workflows. We use \drtulu\citep{shao2025drtulureinforcementlearning} as the backbone agent for all corpus-search experiments. We modify \drtulu's MCP service so it can only use our provided retriever as the search tool. We study three retrievers, which are as follows: BM25~\citep{robertson-1994-okapi}, a spase retriever; gte-Qwen-2-7B-instruct~\citep{li2023towards}, a LLM-based retriever, and ReasonIR~\citep{shao2025reasonir}, a reasoning-intensive retriever. 

\paragraph{Retrieval Index Construction.}
Before experiment, we first download all PDFs according to the URLs in the \ours dataset (detailed in Section~\ref{sec:corpus}). We then convert them to markdown using PyMuPDF~\citep{PyMuPDF} for text and PDFPlumber~\citep{pdfplumber} for tables. Next, we embed the first 32,000 tokens of each markdown file with the corresponding retriever to ensure that the vast majority of each PDF's content is retained while matching the maximum input length of gte-Qwen-2-7B-instruct. We embed each document individually, setting the batch size to 1 to avoid unnecessary padding. Both ReasonIR and gte‑Qwen‑2‑7B-instruct embeddings are computed on a single H100 GPU.
We present the paper length distribution for all four domains in the Appendix \autoref{fig:len_computer_science},  \autoref{fig:len_healthcare}, \autoref{fig:len_humanities} and \autoref{fig:len_natural_science}.

\paragraph{Retrieval Setup.}
During experiments, the \drtulu agent is deployed on two H100 GPUs, where one running vLLM for answer generation and the other running MCP powered by the selected retriever. We set the maximum search iteration to 10 and for each retriever we evaluate two settings for the number of results returned per search, which are top-5 and top-10 (i.e. $k{=}5$ and $k{=}10$). Each retrieval step return a list of paper titles together with their abstracts.

\subsection{Corpus-Search Results}
\label{sec:corpus_search_result}
\begin{figure}[t]
    \centering
    \includegraphics[width=\columnwidth, height=0.4\textheight, keepaspectratio]{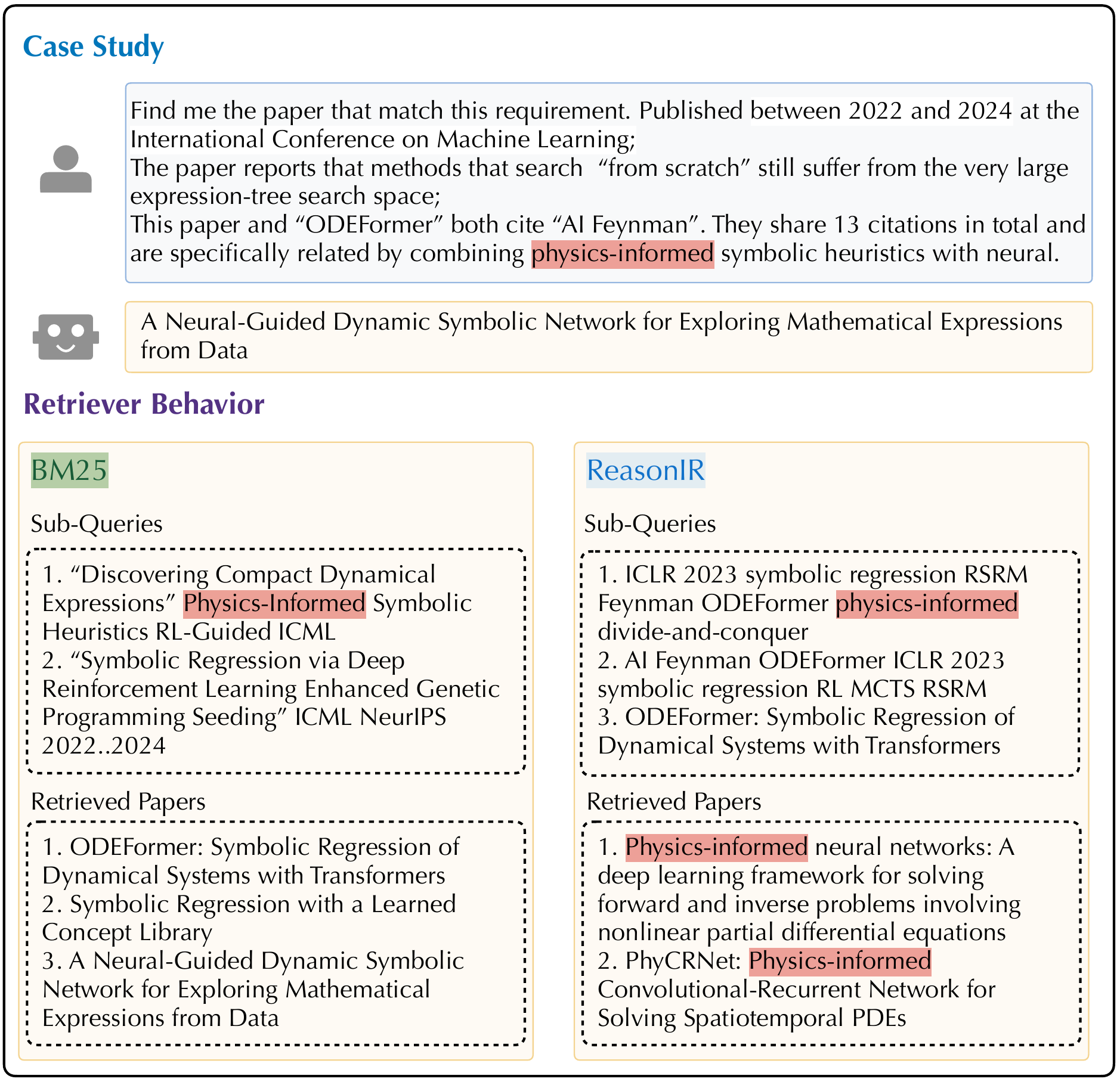}\
    \caption{An illustrative case where LLM-based retrieval fails due to semantic drift. The query seeks a paper that uses physics-informed heuristics. ReasonIR over-emphasizes title-level keywords (highlighted in red) and thus retrieves wrong papers. The retrieved content then reinforces this focus in subsequent retrieval steps, creating a feedback loop that increasingly prioritizes ``physics-informed'' in title. In contrast, BM25 remains anchored by lexical matching in similar sub-queries and avoids this drift.}
    \label{fig:dense_case_study}
\end{figure}
\autoref{tab:Main_Result} presents our results of \drtulu using in-house retrievers as search tools. We have the following main findings:

\paragraph{BM25 dominates LLM-based retrievers on short-form questions, while the gap for open-ended questions is narrower.} On short-form questions, BM25 significantly outperforms LLM-based retrievers by roughly 30\%, suggesting that sparse lexical matching is better aligned with multi-constraint evidence retrieval in this setting.
On open-ended questions, BM25 and gte-Qwen-2-7B-instruct achieve comparable performance, while ReasonIR ranks last on both query types. Notably, gte-Qwen-2-7B-instruct can even slightly outperform BM25, indicating that LLM-based retrieval can be competitive when evaluation tolerates broader evidence coverage. A case for BM25 beating LLM-based retrievers is presented in \autoref{fig:dense_case_study}.

\paragraph{Increasing per-search top-$k$ consistently improves performance.} Across all retrievers, increasing the per-search top-$k$ yields measurable gains, and ReasonIR benefits the most. This suggests that a larger candidate set partially compensates for weaker first-page ranking, especially for LLM-based retrievers.

\paragraph{Query-retriever mismatch limits the value of LLM-based semantics.} A key issue is a pronounced \emph{Query-Retriever Mismatch}: although LLM-based retrievers are trained on natural-language queries, agents often generate keyword-like sub-queries, as shown in Appendix \autoref{fig:Dr_Tulu_decompose}, which poorly match the retrievers’ training distribution and can underutilize semantic capabilities.

\paragraph{LLM-based retrievers suffer from reduced diversity under long-document constraints.} LLM-based retrievers also face information loss when documents approach the maximum input length, and embedding convergence can further reduce per-search diversity. We define \emph{Unique References per Search} (URS) as the average number of retrieved documents returned per search call, computed as the ratio of the average number of documents to the average number of searches.
Under top-5 on short-form questions, BM25 achieves URS of 2.97, whereas ReasonIR attains only URS of 1.98.  This indicates that LLM-based retrievers are less effective to surface the target document under a fixed search budget.

\paragraph{Low-diversity decomposition blunts retriever differences on open-ended queries.} \drtulu exhibits relatively low diversity in its query decomposition. BM25 appears more compatible with DR Tulu’s decomposition and is more robust to long documents, but it does not open a clear advantage on open-ended queries. A plausible explanation is that DR Tulu’s sub-queries cover only a limited portion of the evidence space, so even when retrievers behave differently, multiple ground-truth targets are only partially retrieved.

\begin{table}[t]
\centering
\footnotesize
\setlength{\tabcolsep}{3pt}
\begin{tabular}{lrrrr}
\toprule
\textbf{Method} & \textbf{EM} & \textbf{$\Delta$Met.} & \textbf{$\Delta$Det.} & \textbf{$\Delta$Rel.} \\
\midrule
\multicolumn{5}{l}{\textit{Web Search}} \\
\addlinespace[0.2em]
~~GPT-5 & 71.69 & \cellcolor{red!25}\textbf{-24.45} & -3.85 & -16.92 \\
~~DR Tulu & 42.00 & -17.92 & \cellcolor{red!25}\textbf{-21.60} & -6.88 \\
~~Gemini-2.5-Pro & 38.50 & -11.43 & \cellcolor{red!25}\textbf{-14.69} & -5.67 \\
\midrule
\multicolumn{5}{l}{\textit{Corpus Search}} \\
\addlinespace[0.2em]
~~DR Tulu (BM25) & 75.84 & -8.17 & -14.94 & \cellcolor{red!25}\textbf{-24.59} \\
~~DR Tulu (ReasonIR) & 38.72 & -5.33 & -9.84 & \cellcolor{red!25}\textbf{-15.99} \\
\bottomrule
\end{tabular}
\caption{Ablation study on short-form questions components. EM denotes Exact Match. $\Delta$ denotes the relative accuracy change (\%) when removing each component: metadata (Met.), multimodality detail information (Det.), and relationship constraints (Rel.). \colorbox{red!25}{\textbf{Highlighted}} cells indicate the most impactful component for each method.}
\label{tab:ablation_short_form}
\end{table}
\subsection{Ablation}
\label{sec:ablation}

We conduct ablation studies using short-form questions, as their answers are easier to verify. As discussed earlier, these questions span three aspects of query information: paper metadata, multimodal details, and inter-paper relationships. Manual inspection shows that leveraging any two of these components is sufficient to locate 93.67\% of the target papers. Based on this observation, we examine how deep research agents exploit different sources of query information. For each model family, we select one model and report results in \autoref{tab:ablation_short_form}.

\paragraph{Search method strongly shapes which information matters.} Different deep-research agents emphasize different components of the query, and this emphasis shifts with the search method. Under web search, DR Tulu is most sensitive to paper details, whereas under corpus-based search, inter-paper relationships become the dominant factor. Moreover, agents that share the same search method exhibit similar sensitivity patterns. For instance, both DR Tulu and Gemini-2.5-Pro rely on Google Search and are most influenced by paper details, indicating that the retrieval backend largely determines which part of query information drive performance.

\section{Test-Time Corpus Scaling}
Our analysis in Section~\ref{sec:experiments} reveals a fundamental limitation of existing deep research agents: certain papers requiring intensive reasoning are inherently difficult to retrieve. Prior work~\cite{bright2024, shao2025reasonir} proposes to address this challenge through test-time scaling on the query side, augmenting queries with reasoning chains. In contrast, we propose an alternative form of test-time scaling at the document corpus side. The key intuition is that, rather than increasing query complexity, we incorporate reasoning-derived information into documents, making them easier to retrieve for off-the-shelf retrievers. 

\subsection{Method}
Since \drtulu primarily issues keyword-based queries, we augment each document’s Markdown by prepending salient keywords to improve retrieval effectiveness. Specifically, we first obtain key bibliographic metadata, including publication venue, year, authors, and citation counts.  In addition, we use Qwen3-Next-80B-A3B-Instruct~\citep{qwen_2025_next} to process the Markdown and extract eight topic-relevant keywords that summarize the paper’s core contributions. These fields are formatted as emphasized keywords and prepended to each document, so that both bibliographic signals and high-level semantic cues are surfaced for effective keyword-based retrieval.\footnote{We scale the corpus by augmenting documents with additional information in bag of keywords. With LLMs, future work could explore more aggressive corpus scaling strategies, such as directly editing or rewriting each paper.}

\begin{table}[t]
\centering
\footnotesize
\setlength{\tabcolsep}{3pt}
\begin{tabular}{l cc c cc}
\toprule
& \multicolumn{2}{c}{\textbf{Short-form}} & & \multicolumn{2}{c}{\textbf{Open-ended}} \\
\cmidrule{2-3} \cmidrule{5-6}
\textbf{Retriever} & Before & After & & Before & After \\
\midrule
BM25 & 75.80 & 83.98 \colorbox{green!20}{\tiny+8.18} & & 25.52 & 27.25 \colorbox{green!20}{\tiny+1.73} \\
gte-Qwen & 46.90 & 47.80 \colorbox{green!20}{\tiny+0.90} & & 26.03 & 27.82 \colorbox{green!20}{\tiny+1.79} \\
ReasonIR & 38.70 & 40.40 \colorbox{green!20}{\tiny+1.70} & & 17.25 & 19.79 \colorbox{green!20}{\tiny+2.54} \\
\bottomrule
\end{tabular}
\caption{Performance before and after corpus-level test-time scaling. Short-form is evaluated by Exact Match (EM) (\%) and open-ended by Weighted Recall (\%). Improvements are shown with \colorbox{green!20}{green background}.}
\label{tab:corpus_scaling}
\end{table}

\subsection{Results}
In this experiment, we set the maximum number of search iterations to 10 and retrieve the top-5 results per search. \autoref{tab:corpus_scaling} reports the results of \drtulu with three different retrievers, both before and after applying test-time corpus scaling.

\paragraph{BM25 benefits most from corpus scaling.} On short-form questions, BM25 achieves absolute gain of 8.18\%, LLM-based retrievers exhibit only modest improvements. This is largely because BM25 is more sensitive to keyword signals, while LLM-based retrievers, as discussed, struggle when documents approach input‑length limits. Therefore, the added information makes documents only marginally easier for them.

\paragraph{Limited improvement on open‑ended questions.} All three retrievers show only marginal improvements on open-ended questions. This result aligns with our earlier observation at section \ref{sec:corpus_search_result} that \drtulu’s (and other deep research agents) generated query lacks diversity, which limits retrieval breadth and prevents corpus-level scaling from fully translating into downstream performance gains.


\section{Conclusion}
We introduce \ours, a benchmark for reasoning-intensive scientific literature retrieval. Through extensive evaluation, we reveal a critical finding: LLM-based retrievers underperform BM25 by approximately 30\% in deep research agent workflows, as existing agents generate keyword-oriented sub-queries. To address this limitation, we propose corpus-level test-time scaling, which enriches papers with metadata and LLM-generated keywords, and achieves consistent improvements. Our work highlights that effective collaboration between retrievers and agents requires further adaptation.

\section*{Limitations and Future Work}
We acknowledge limitations in our study. We do not perform instruction fine-tuning or alignment on the open-source deep-research agents. As a result, we are unable to assess whether training agents to adapt their query generation strategies based on the underlying retriever type could improve performance. Exploring such retriever-aware agent training remains a valuable direction for future work. Additionally, most of our behavioral analysis is conducted on DR Tulu, whose post-training procedures may significantly influence the observed agent behaviors. Consequently, our findings may not fully generalize to agents with different training recipes or base model architectures.

\section*{Acknowledgements}

Tiansheng Hu and Chen Zhao were supported by NYU Shanghai Center for Data Science. This work was supported in part through the NYU IT High Performance Computing resources, services, and staff expertise.
\bibliography{custom}

\newpage
\appendix

\addtocontents{toc}{\protect\setcounter{tocdepth}{3}}
\renewcommand{\contentsname}{\large Appendix Contents}
\hypersetup{linkcolor=black}
\tableofcontents

\newpage
\section{Appendix}
\subsection{Query-Answer Example}
\begin{figure}[!htbp]
\centering
\begin{tcolorbox}[colback=green!5, colframe=green!70!black, boxrule=0.8pt, arc=3pt, left=5pt, right=5pt, top=5pt, bottom=5pt]

\textbf{Query:}

\vspace{2pt}
\small
Find me the paper that matches this requirement:
\begin{itemize}[leftmargin=*, itemsep=1pt, topsep=2pt]
    \item Published between 2023 and 2025 (inclusive), presented at the Annual Meeting of the Association for Computational Linguistics, and building on 40 prior works.
    \item A key visualization shows their method dramatically improves T5-small versus the vanilla baseline: typical JGA gains of ~0.15–0.25, with a striking +0.25 jump on Task 4 (their method T5 = 0.58 vs vanilla T5 = 0.33).
    \item This paper and ``APT: Adaptive Pruning and Tuning Pretrained Language Models for Efficient Training and Inference'' both cite ``Movement Pruning: Adaptive Sparsity by Fine-Tuning'' for weight-gradient salience scoring to guide which parameters to prune or tune, and ``Adaptive Budget Allocation for Parameter-Efficient Fine-Tuning'' for sensitivity, smoothing, and uncertainty measures to allocate and remove tuning parameters. They share 7 citations in total and are specifically related by salience-driven pruning and adaptive budget allocation for parameter-efficient fine-tuning.
\end{itemize}

\vspace{4pt}
\hrule
\vspace{4pt}

\textbf{Answer:}

\vspace{2pt}
\small
TaSL: Continual Dialog State Tracking via Task Skill Localization and Consolidation

\end{tcolorbox}
\caption{Example of a Short-Form question.}
\label{fig:browsecomp_example}
\end{figure}
\begin{figure}[!htbp]
\centering
\begin{tcolorbox}[colback=green!5, colframe=green!70!black, boxrule=0.8pt, arc=3pt, left=5pt, right=5pt, top=5pt, bottom=5pt]

\textbf{Query:}

\vspace{2pt}
\small
I am working on multi-agent reinforcement learning in general-sum Markov games, specifically focusing on learning Nash Equilibrium, Coarse Correlated Equilibrium, and Correlated Equilibrium using oracle-based algorithmic frameworks.

\vspace{3pt}
I am considering comparing sample-efficient methods for equilibrium learning under stationary environments with general function approximation (as in MAMEX and related complexity measures) against black-box, multi-scale restart approaches designed for non-stationary, time-varying settings.

\vspace{3pt}
Please give me some background and methodological basis references for establishing theoretical foundations and performance benchmarks for regret minimization and equilibrium computation in MARL.

\vspace{4pt}
\hrule
\vspace{4pt}

\textbf{Answer:}

\vspace{2pt}
\small
\begin{itemize}[leftmargin=*, itemsep=1pt, topsep=2pt]
    \item A Black-box Approach for Non-stationary Multi-agent Reinforcement Learning
    \item Sample-Efficient Multi-Agent RL: An Optimization Perspective
    \item ...
\end{itemize}

\end{tcolorbox}
\caption{Example of a Open-Ended question.}
\label{fig:natural_example}
\end{figure}
\subsection{Query-Decomposition Case Study}
\begin{figure}[!htbp]
\centering
\begin{tcolorbox}[colback=blue!5, colframe=blue!70!black, boxrule=0.8pt, arc=3pt, left=5pt, right=5pt, top=5pt, bottom=5pt]

\textbf{Query:}

\vspace{2pt}
\small
Find me the paper that matches this requirement:
\begin{itemize}[leftmargin=*, itemsep=1pt, topsep=2pt]
    \item Published in the period from 2021 to 2025, presented at the Annual Meeting of the Association for Computational Linguistics conference, and written by 5 authors.
    \item Its primary figure demonstrates that the architecture explicitly leverages both source KBQA training data and target KBQA few-shot examples to influence retrieval, reranking, and generation, highlighting transfer from source domains plus few-shot adaptation to the target as a core design choice.
    \item This paper and ``RetinaQA: A Robust Knowledge Base Question Answering Model for both Answerable and Unanswerable Questions'' both cite ``Few-shot In-context Learning on Knowledge Base Question Answering'' for its few-shot in-context generate-then-ground sketching pipeline and ``Beyond I.I.D.: Three Levels of Generalization for Question Answering on Knowledge Bases'' for the GrailQA dataset and s-expression logical-form formalism. They share 17 citations in total and are specifically related by emphasis on transferable sketch-based logical-form generation and evaluation on GrailQA.
\end{itemize}

\vspace{4pt}
\hrule
\vspace{4pt}

\textbf{Decomposed Search Queries:}

\vspace{2pt}
\small
\begin{enumerate}[leftmargin=*, itemsep=1pt, topsep=2pt]
    \item \texttt{site:aclanthology.org ACL 2021 knowledge base question answering GrailQA sketch generation transfer few-shot}
    \item \texttt{site:aclanthology.org 2024.acl knowledge base question answering few-shot transfer GrailQA sketch}
    \item \texttt{RetinaQA arXiv PDF 2403.10849}
    \item \texttt{Semantic Scholar RetinaQA 2403.10849 references}
    \item \texttt{site:semanticscholar.org RetinaQA 2403.10849 references GrailQA}
    \item \texttt{site:aclanthology.org RetinaQA 2024.acl-long}
    \item \texttt{Shu and Yu 2024 KBQA few-shot in-context learning TIARA retriever ACL}
    \item \texttt{ACL 2024 KBQA transfer few-shot retriever reranking generation figure source target few shots site:aclanthology.org 2024.acl-long}
    \item \texttt{Beyond I.I.D.: Three Levels of Generalization for Question Answering on Knowledge Bases ACL Anthology 2021}
\end{enumerate}

\vspace{4pt}
\hrule
\vspace{4pt}

\textbf{Answer:}

\vspace{2pt}
\small
Few-shot Transfer Learning for Knowledge Base Question Answering: Fusing Supervised Models with In-Context Learning

\end{tcolorbox}
\caption{GPT-5 Query Decomposition Example.}
\label{fig:gpt_5_decompose}
\end{figure}
\subsection{Document Length Distribution}
\begin{figure}[!htbp]
\centering
\begin{tcolorbox}[colback=blue!5, colframe=blue!70!black, boxrule=0.8pt, arc=3pt, left=5pt, right=5pt, top=5pt, bottom=5pt]

\textbf{Query:}

\vspace{2pt}
\small
Find me the paper that matches this requirement:
\begin{itemize}[leftmargin=*, itemsep=1pt, topsep=2pt]
    \item Published in the period from 2021 to 2025, presented at the Annual Meeting of the Association for Computational Linguistics conference, and written by 5 authors.
    \item Its primary figure demonstrates that the architecture explicitly leverages both source KBQA training data and target KBQA few-shot examples to influence retrieval, reranking, and generation, highlighting transfer from source domains plus few-shot adaptation to the target as a core design choice.
    \item This paper and ``RetinaQA: A Robust Knowledge Base Question Answering Model for both Answerable and Unanswerable Questions'' both cite ``Few-shot In-context Learning on Knowledge Base Question Answering'' for its few-shot in-context generate-then-ground sketching pipeline and ``Beyond I.I.D.: Three Levels of Generalization for Question Answering on Knowledge Bases'' for the GrailQA dataset and s-expression logical-form formalism. They share 17 citations in total and are specifically related by emphasis on transferable sketch-based logical-form generation and evaluation on GrailQA.
\end{itemize}

\vspace{4pt}
\hrule
\vspace{4pt}

\textbf{Decomposed Search Queries:}

\vspace{2pt}
\small
\begin{enumerate}[leftmargin=*, itemsep=1pt, topsep=2pt]
    \item \texttt{RetinaQA knowledge base question answering ACL}
    \item \texttt{FuSIC-KBQA ACL few-shot transfer learning KBQA figure retrieval reranking LLM}
    \item \texttt{RetinaQA ACL 2024 authors GrailQA s-expression citations "Few-shot In-context Learning on Knowledge Base Question Answering"}
    \item \texttt{RetinaQA ACL authors arXiv 2403.10849 GrailQA s-expression}
    \item \texttt{RetinaQA ACL 2024 authors arXiv 2403.10849 GrailQA s-expression citations}
\end{enumerate}

\vspace{4pt}
\hrule
\vspace{4pt}

\textbf{Answer:}

\vspace{2pt}
\small
Few-shot Transfer Learning for Knowledge Base Question Answering: Fusing Supervised Models with In-Context Learning

\end{tcolorbox}
\caption{Dr-Tulu Query Decomposition Example.}
\label{fig:Dr_Tulu_decompose}
\end{figure}

\begin{figure}[!htbp]
    \centering
    \includegraphics[width=\columnwidth, height=0.4\textheight, keepaspectratio]{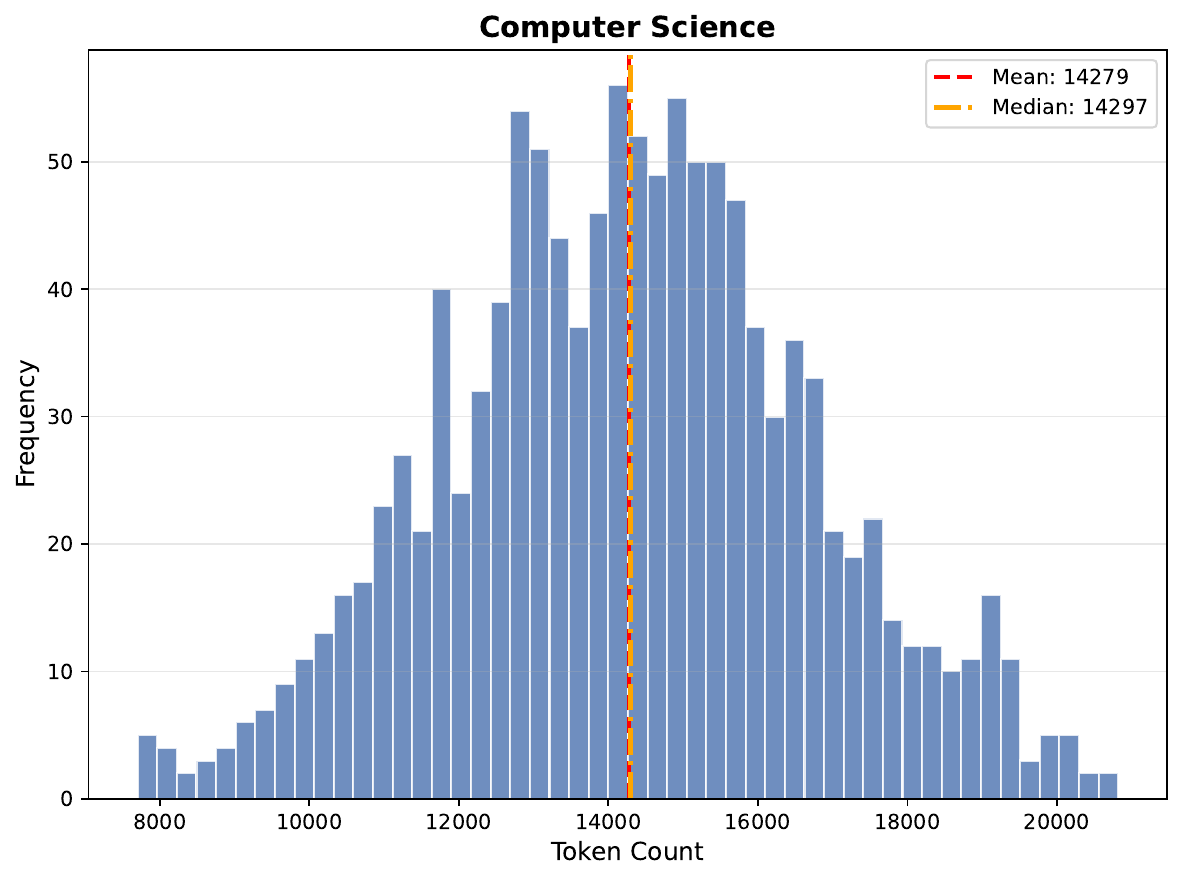}
    \caption{Distribution of markdown length (in tokens) for 1,000 randomly sampled documents from the Computer Science domain.}
    \label{fig:len_computer_science}
\end{figure}
\begin{figure}[!htbp]
    \centering
    \includegraphics[width=\columnwidth, height=0.4\textheight, keepaspectratio]{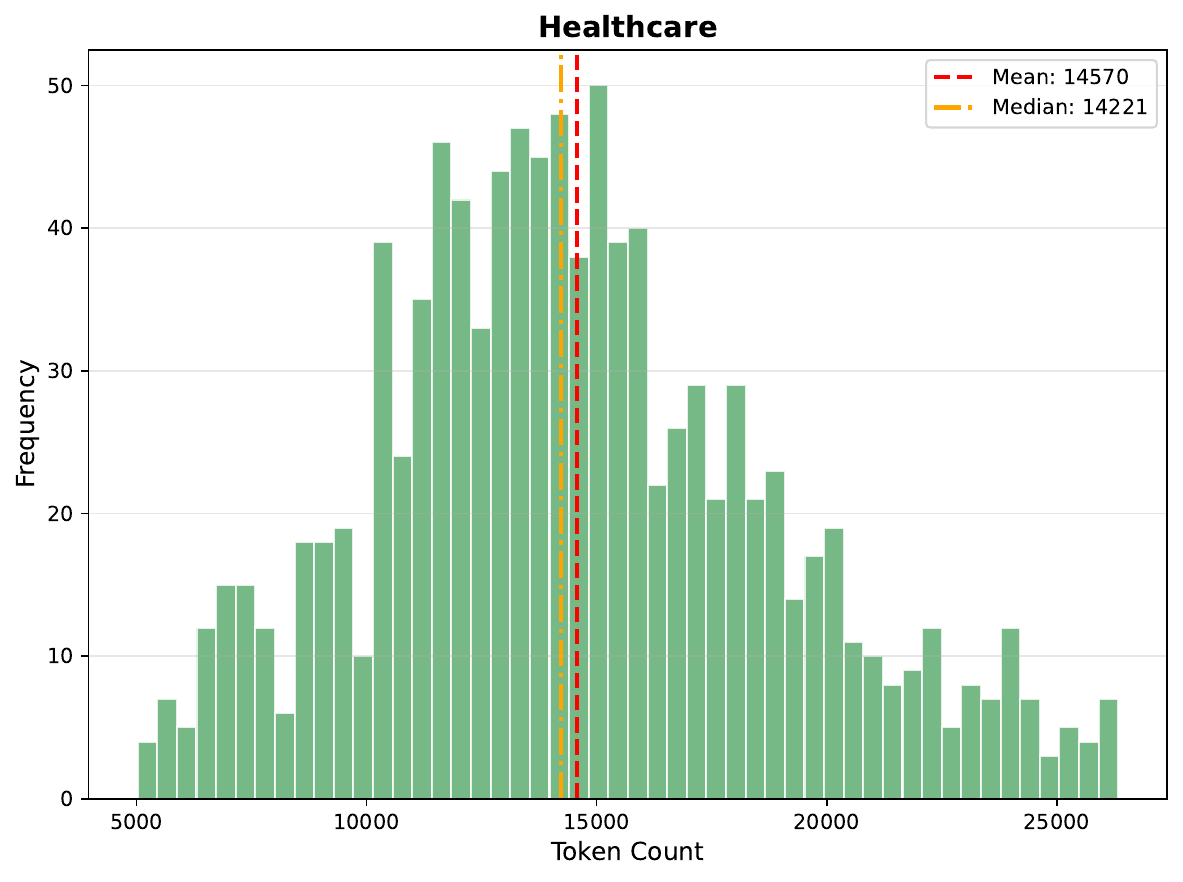}
    \caption{Distribution of markdown length (in tokens) for 1,000 randomly sampled documents from the Healthcare domain.}
    \label{fig:len_healthcare}
\end{figure}
\begin{figure}[!htbp]
    \centering
    \includegraphics[width=\columnwidth, height=0.4\textheight, keepaspectratio]{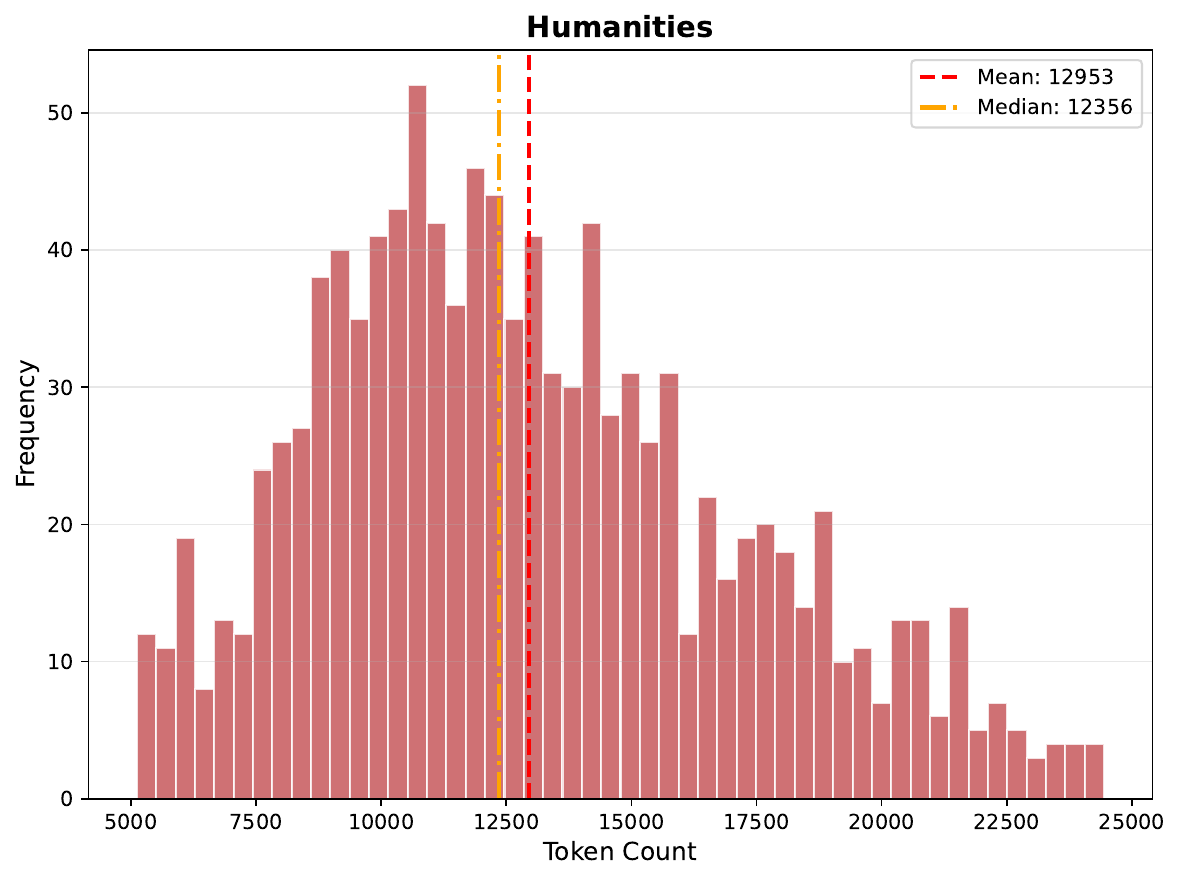}
    \caption{Distribution of markdown length (in tokens) for 1,000 randomly sampled documents from the Humanities domain.}
    \label{fig:len_humanities}
\end{figure}
\begin{figure}[!htbp]
    \centering
    \includegraphics[width=\columnwidth, height=0.4\textheight, keepaspectratio]{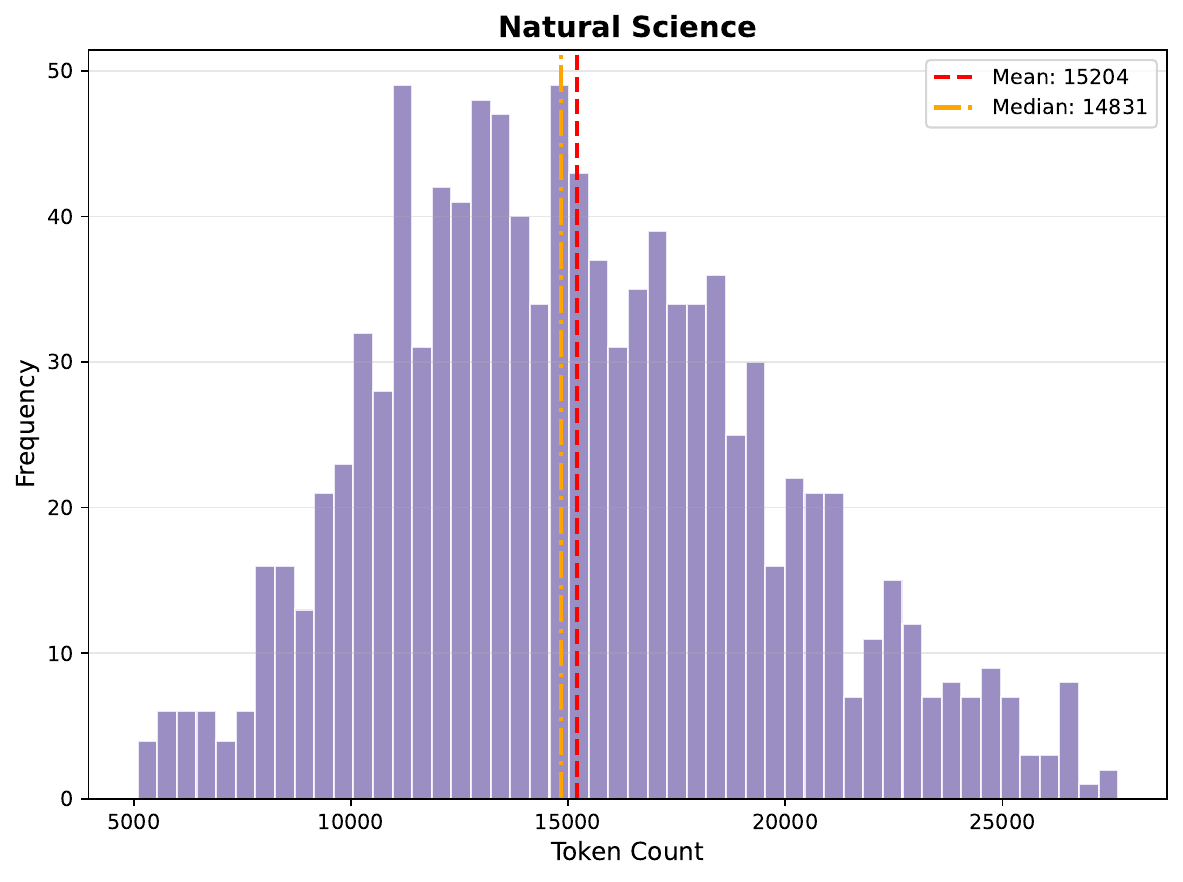}
    \caption{Distribution of markdown length (in tokens) for 1,000 randomly sampled documents from the Natural Science domain.}
    \label{fig:len_natural_science}
\end{figure}

\subsection{Comparison with BrowseComp-Plus: Retriever Behavior}
We observe a retriever ranking that differs from BrowseComp-Plus \citep{chen2025browsecompplusfairtransparentevaluation}. In our experiments, BM25 consistently outperforms LLM-based dense retrievers (e.g., gte-Qwen2-7B-Instruct), whereas BrowseComp-Plus reports stronger performance from dense retrievers such as Qwen3-Embed-8B. We attribute the discrepancy to differences in (i) task characteristics, (ii) retriever implementations, and (iii) agent model strength and query decomposition.

\paragraph{Longer documents and weaker answer locality in our setting.}
BrowseComp-Plus uses substantially shorter documents on average (6733 tokens vs.\ 13376 in ours) and exhibits strong \emph{early-answer locality}: truncating documents to the first 512 tokens still preserves the ground-truth answer in at least one gold document for 86.5\% of queries. This property favors dense retrievers that primarily represent the document prefix. In contrast, our documents are longer and evidence is more dispersed, reducing the effectiveness of limited-window dense encoding.

\paragraph{Asymmetric text coverage can favor dense retrievers under early-answer locality.}
BrowseComp-Plus encodes only the first 4096 tokens for Qwen3-Embed-8B, while BM25 indexes the full document. When answers are front-loaded, prefix-only dense encoding can act as an implicit denoising mechanism, whereas full-text BM25 may incur additional lexical noise. This asymmetric coverage therefore biases the comparison toward dense retrievers. In our experiments, we allow up to 32{,}000 tokens per document, so dense retrievers do not benefit from short-prefix encoding.

\paragraph{Agent strength and query decomposition modulate retriever sensitivity.}
BrowseComp-Plus further suggests that stronger agent models (e.g., GPT-5 and o3) are less sensitive to retriever choice, as the gap between BM25 and Qwen3-Embed-8B narrows compared to weaker models (e.g., Qwen3-32B and Gemini-2.5 Flash). Moreover, BrowseComp-Plus adopts a ReAct-style framework \citep{yao2023react} that produces natural-language sub-queries, while our setup (GPT-5 series, Gemini-2.5 series, and DR-Tulu) uses more keyword-oriented decomposition. This difference in query formulation can shift the relative advantage between lexical and dense retrievers.

\subsection{Further Experiments with SearchR1-32B}
As a supplement to our main experiments with \drtulu, we evaluate another open-source deep-research agent, SearchR1-32B \citep{jin2025searchr}. Table~\ref{tab:SearchR1_Corpus} summarizes corpus-search performance across domains.

\paragraph{SearchR1-32B exhibits near single-shot retrieval.}
SearchR1-32B issues only 1.1--1.2 searches per question, leaving limited room for iterative query refinement. Consequently, end-to-end performance is primarily determined by the initial query formulation and the base retriever.

\paragraph{Natural-language querying does not obviate lexical matching.}
Although SearchR1-32B produces natural-language queries rather than keyword-style decompositions, BM25 remains markedly stronger on short-form questions. On open-ended questions, BM25 and gte-Qwen are closer in performance, while ReasonIR remains substantially worse, consistent with previous findings. Importantly, the average number of references is similar across retrievers, suggesting that the observed differences are driven by retrieval quality rather than references counts.
\begin{table*}[htbp]
\centering
\small
\begin{tabular}{l cccc cc c}
\toprule
\textbf{Method} & \textbf{Com. Sci.} & \textbf{Healthcare} & \textbf{Humanities} & \textbf{Nat. Sci.} & \textbf{Avg. Searches} & \textbf{Avg. Refs} & \textbf{Avg. Perf.} \\
\midrule\midrule
\multicolumn{8}{l}{\colorbox{gray!25}{\textbf{\textsc{~~Corpus Search (SearchR1-32B)~~}}}} \\
\midrule
\multicolumn{8}{l}{\footnotesize\textbf{Short-Form Questions (Exact Match)}} \\[0.1em]
BM25  $k{=}5$     & 19.57 & 45.26 & 34.13 & 40.91 & 1.21 & 5.7 & 34.97 \\
gte-Qwen  $k{=}5$      & 4.29 & 23.44 & 26.23 & 16.18 & 1.24 & 5.7 & 17.54 \\
ReasonIR  $k{=}5$ & 3.79 & 20.90 & 15.70 & 11.28 & 1.21 & 5.5 & 12.92 \\
\addlinespace[0.5em]
\multicolumn{8}{l}{\footnotesize\textbf{Open-Ended Questions (Weighted Recall)}} \\[0.1em]
BM25  $k{=}5$     & 6.80 & 12.58 & 13.52 & 10.20 & 1.16 & 5.6 & 10.77 \\
gte-Qwen $k{=}5$      & 7.57 & 11.78 & 12.73 & 7.48 & 1.10 & 5.2 & 9.89 \\
ReasonIR  $k{=}5$ & 4.65 & 5.52 & 8.19 & 2.93 & 1.11 & 5.4 & 5.32 \\
\bottomrule
\end{tabular}
\caption{Corpus-search results with SearchR1-32B.}
\label{tab:SearchR1_Corpus}
\end{table*}

\subsection{Prompt Templates}
\begin{figure*}[t]
\begin{tcolorbox}[colback=black!7.5!white, colframe=black!40!black, title=Academic Paper Keyword Generation Prompt, 
fontupper=\footnotesize, fonttitle=\footnotesize]
Based on the following academic paper content, generate exactly 8 keywords that best represent the main topics, methods, or contributions of this paper.

Content:
\{content[:20000]\}

Return ONLY the 8 keywords separated by commas, nothing else. Example format:
keyword1, keyword2, keyword3, keyword4, keyword5, keyword6, keyword7, keyword8
\end{tcolorbox}
\caption{Prompt of Academic Paper Keyword Generation}
\label{Keyword Prompt}
\end{figure*}
\begin{figure*}[t]
\begin{tcolorbox}[colback=black!7.5!white, colframe=black!40!black, title=Prompt for Analyzing Shared Reference Functions, 
fontupper=\footnotesize, fonttitle=\footnotesize]
\textbf{System:} You are a research assistant helping to analyze academic citations. Provide concise, accurate summaries.

\vspace{0.5em}
\textbf{User:}

Shared Citation Paper: "\{shared\_title\}" by \{shared\_authors\}

Paper 1: Target Paper \\
- Citing Contexts: \\
\hspace*{1em} 1. \{context\_1\} \\
\hspace*{1em} 2. \{context\_2\} \\
\hspace*{1em} ... \\
- Intents: \{intents\}

Paper 2 (shared \{shared\_count\} papers with Target Paper): "\{cited\_paper\_title\}" \\
- Citing Contexts: \\
\hspace*{1em} 1. \{context\_1\} \\
\hspace*{1em} 2. \{context\_2\} \\
\hspace*{1em} ... \\
- Intents: \{intents\}

Task: Summarize in ONE sentence what role this shared citation paper played for both papers. Focus on the specific contributions or methods it provided to each paper.
\end{tcolorbox}
\caption{Prompt for Analyzing the Functional Role of Shared References Between Two Papers}
\label{fig:shared_ref_prompt}
\end{figure*}
\begin{figure*}[t]
\begin{tcolorbox}[colback=black!7.5!white, colframe=black!40!black, title=Prompt for Generating Comprehensive Summaries of Paper Relationships, 
fontupper=\footnotesize, fonttitle=\footnotesize]
\textbf{System:} You are a research assistant analyzing paper relationships. Write concise, specific summaries:
\begin{itemize}[noitemsep, topsep=0pt, leftmargin=1em]
    \item Maximum 2 sentences
    \item Mention ONLY 2 most important shared citations by title
    \item Use `This paper' for Paper 1, full title for Paper 2
    \item Be specific about what the citations are used for
    \item Avoid generic phrases and long lists
\end{itemize}

\vspace{0.5em}
\textbf{User:}

Paper 1 (This Paper): ``\{paper1\_title\}'' \\
Paper 2: ``\{cited\_paper\_title\}''

These two papers share \{shared\_count\} common citations.

\textbf{Analysis of shared citations:}

1. ``\{citation\_title\}'' by \{citation\_authors\} \\
\hspace*{1em} Paper 1 uses it for: \{intents\} \\
\hspace*{1em} Paper 2 uses it for: \{intents\} \\
\hspace*{1em} How both papers use it: \{summary\}

2. ... (additional shared citations)

\textbf{Task:} Write a concise summary (2 sentences maximum) that:
\begin{enumerate}[noitemsep, topsep=0pt, leftmargin=1.5em]
    \item Uses `This paper' to refer to Paper 1
    \item Uses the full title when referring to Paper 2
    \item Selects and mentions 2 important shared citations by title (in quotes). Avoid famous papers mentioned only as convention, e.g., `Attention is All You Need'
    \item Briefly explains what commonality these key citations reveal
    \item Mentions that they share \{shared\_count\} citations in total
    \item Be specific and concrete---avoid generic statements
\end{enumerate}

\textbf{Format:}
\begin{itemize}[noitemsep, topsep=0pt, leftmargin=1em]
    \item First sentence: Introduce the 2 key shared citations and their specific role
    \item Second sentence: State ``They share \{N\} citations in total and are specifically related by [brief commonality].''
\end{itemize}

\textbf{Additional constraints:} No parentheses or brackets; keep each part simple and direct.
\end{tcolorbox}
\caption{Prompt for Generating Comprehensive Summaries of Shared References Relationships Between Paper Pairs}
\label{fig:comprehensive_summary_prompt}
\end{figure*}
\begin{figure*}[t]
\begin{tcolorbox}[colback=black!7.5!white, colframe=black!40!black, title=Prompt for Selecting the Most Characteristic Summary, 
fontupper=\footnotesize, fonttitle=\footnotesize]
\textbf{System:} You are a helpful assistant that selects the most characteristic summary. Return only a single integer.

\vspace{0.5em}
\textbf{User:}

You are given a list of paper summaries. Each summary describes the relationship between a source paper and a cited paper.

\textbf{Your task:} Select exactly 1 summary that is MOST characteristic and informative.
\begin{itemize}[noitemsep, topsep=0pt, leftmargin=1em]
    \item Choose the summary that provides the most specific, concrete technical details
    \item Prefer summaries that mention distinctive methods, techniques, or research approaches
    \item Avoid generic or vague descriptions
\end{itemize}

\textbf{Summaries:}

[0] \{summary\_0\}

[1] \{summary\_1\}

[2] \{summary\_2\}

... (additional summaries)

Return ONLY a single integer index (e.g., 0, 1, 2, etc.) \\
No explanation needed, just the number.

\vspace{0.8em}
\tcbline
\vspace{0.3em}

\textbf{Query Construction:}

After selecting index $i$, the final query is constructed as:

\texttt{Find me the paper that have the following characteristics: \{selected\_summary\}}

\end{tcolorbox}
\caption{Prompt for Selecting the Most Characteristic Summary from Multiple Paper Relationship Descriptions and Constructing a Retrieval Query}
\label{fig:characteristic_selection_prompt}
\end{figure*}

\end{document}